\documentclass[preprint,aps,showpacs]{revtex4}
\usepackage{graphicx}
\usepackage{bm}
\begin{document}
\title{Is SU(2) lattice gauge theory a spin glass?}
\author{Michael Grady}
\email{grady@fredonia.edu}
\affiliation{Department of Physics\\ State University of New York at Fredonia\\
Fredonia NY 14063 USA}
\date{\today}
\thispagestyle{empty}
\begin{abstract}
A new order parameter is constructed for SU(2) lattice gauge theory 
in the context of the two-real-replica method normally used for spin glasses.  The order parameter is 
sensitive to a global Z2 subgroup of the gauge symmetry which is seen to break 
spontaneously at $\beta = 4/g^2 = 1.96\pm 0.01$. No gauge fixing is required. Finite size scaling is 
consistent with a high-order paramagnet
to spin glass transition with a critical exponent $\nu = 0.99 \pm 0.13$. The existence of this 
transition suggests a second transition from spin glass 
to ferromagnet should exist at higher $\beta$.

\end{abstract}
\maketitle
\section{Introduction}
It has been recognized for some time that spin glasses and gauge theories have similarities. In 
fact many spin glasses have hidden gauge symmetries.  The motivation for treating the 4-d SU(2) 
lattice gauge theory as a possible spin glass arises from studying the theory in minimal
Coulomb gauge; however, the spin-glass techniques used in this work enable one to entirely
avoid any gauge fixing.  Because iterative gauge fixing schemes are cumbersome and add possible systematic errors,
avoiding gauge fixing when possible leads to a cleaner and more robust result. Once the gauge theory is reformulated as an
annealed spin glass as described below, a standard two-real-replica simulation is used to detect the possible
existence of spin-glass order.  Using finite size scaling methods, a phase transition from a paramagnet 
to a spin glass is found in 3-d hyperlayers of the 4-d theory 
at $\beta=4/g^2 \simeq 1.96$, close to or coincident with the previously known roughening transition. This transition 
breaks a Z2 subgroup of the gauge symmetry, global in three dimensions and local in one.  Not only is this interesting for 
understanding lattice gauge theory,
but it also gives a new example of a 3-d spin glass which seems relatively easy to study by Monte Carlo (MC) methods.  
For instance,
the Binder 4th-order cumulant shows a very clear crossing, which is not always the case for spin glasses.  

\section{Coulomb Gauge Magnetization}
The minimal Coulomb gauge attempts to maximize the traces of all links lying in three of the four 
Euclidean spatial directions.  At weak coupling, this results in links quite close to the identity, on average.
For instance, at $\beta =2.9$ the average link trace is $\approx 0.927$.  At weaker $\beta$ it appears to
closely track the 
fourth root of the plaquette\cite{me1,me2}. In this gauge, the fourth-dimension pointing links can be interpreted as
magnetic spins, with their spatial average over perpendicular hyperlayers as magnetizations.
A global SU(2) remnant gauge symmetry still exists on each hyperlayer, because such a transition leaves the
traces of sideways links within the hyperlayer invariant, and thus does not disturb the gauge condition. 
In other words, if a sideways link is written as
\[
U = a_0 \protect{\bf{1}} + i\sum _{j=1}^{3} a_j \tau _j
\]
the $a_0$ component is not affected, which is what the gauge condition is attempting to maximize.
To further engage the spin-model analogy, the fourth-dimension pointing links are most
easily thought of as unit vectors in a four-dimensional O(4) spin 
space $\vec{s} = (a_0$, $a_1$, $a_2$, $a_3$). Since these are transformed by remnant SU(2) transformations on both
adjacent hyperlayers the full symmetry space of each layer treated individually is SU(2)$\times$SU(2)$=$O(4). 
The utility of these spin-like variables definable in the Coulomb gauge, and the connection of remnant
symmetry breaking with deconfinement has been recognized for a long time\cite{mari}. However,
such methods did not become popular, most likely due to the practical difficulties of iterative gauge-fixing.

\section{Spin-Glass Formulation}
In the weak 
coupling limit, $\beta \rightarrow \infty$, when the gauge condition will be able to set the sideways links 
closer and closer to the identity, the plaquette interaction collapses to become a 
nearest-neighbor scalar product between the O(4) spins, and the 
theory becomes a stack of non-interacting 3-d O(4) Heisenberg models. This dimensional reduction occurs
because, unlike spin theories, in a gauge theory a 4th-dimension link interacts directly 
with other 4th-dimension links which
are displaced from it in {\em perpendicular directions only} (what we are calling sideways directions).  
Away from $\beta = \infty$ 
where  the sideways links are not quite the identity, the theory can still be considered a set of 3-d
O(4) Heisenberg models but with
additional interactions that are small, or occasionally large, but rare.  These interactions couple the layers, 
though not directly, and they vary spatially.  This picture of the sideways links as being essentially
parameters controlling additional interactions in an otherwise ferromagnetic spin model leads to the interpretation 
as a spin glass.  The MC is pictured as nested.  First, an ordinary 4-d MC is run
awhile. Then the sideways links are fixed and treated as ``disorder interactions" of a spin glass, with the 4th 
dimension pointing links as the spins. These are subjected themselves to a second MC, which explores how 
the ``magnet" behaves in that particular disorder environment.  The full behavior is determined by nested averages,
first over the inner magnetic MC and then over the outer ``disorder" MC.
If this is all that is done it is no more than an odd order of doing an ordinary MC simulation. However, if 
two replicas are spawned for each inner MC, then correlations between them can yield new information.  In
spin glasses one cannot average the magnetization spatially, because the disordering interactions destroy the
spontaneous magnetization.  However there are spatially-varying hidden patterns of magnetization that can be teased out
by either studying local spins in a given replica over MC time (Edwards-Anderson order parameter) or by correlating
the two independent replicas.  If the two replicas remain correlated over long times, then ergodicity is broken in the infinite
lattice and one has a spin-glass phase with a non-zero spin-glass magnetization. Of course, for 
low enough disorder and temperature, 
the spin glass normally enters a
ferromagnetic phase, where a conventional spontaneous magnetization also arises.  At high temperatures it 
enters a paramagnetic 
fully-ergodic phase where both the spin glass and ordinary magnetizations vanish.

For most spin glasses, there are separate parameters for temperature and disorder. For low temperatures, the system
is a ferromagnet for low interaction disorder, and has a transition to the spin-glass phase at higher disorder. 
For instance, a 3-d
O(3) Heisenberg $\pm J$ spin glass has a spin-glass transition at zero temperature when 21\% of randomly chosen 
interactions are switched from 
ferromagnetic to anti-ferromagnetic\cite{3dhsg}. Both
low-temperature phases, the ferromagnet and the spin glass, 
have thermal transitions to a paramagnetic phase at high temperatures.  If the lattice gauge theory is to be interpreted 
as a spin glass it needs to be recognized that here a single parameter, $\beta$, controls both the 
pseudo-temperature ($T=1/ \beta$) and the 
degree of disorder induced by the deviation of sideways links from the identity.  Thus it is in the category of an
annealed spin glass as opposed to the more usual quenched spin glass, since the disordering interactions are themselves
chosen from a thermal ensemble.  Due to the apparent robustness
of ferromagnetic order in the 3-d Heisenberg model with respect to the addition of small amounts of other interactions 
(such as making a considerable percentage of the links antiferromagnetic), it seems 
reasonable that the ferromagnetic phase which must exist at 
$\beta= \infty$ (where the identification with layered Heisenberg models is exact) will continue to exist at 
large but finite
$\beta$. Of course this is controversial, since it has been assumed that the non-abelian lattice gauge theories have
only one phase for all $\beta$, and that the continuum limit is confining. It has been 
shown that a {\em ferromagnetic} phase in
minimal Coulomb gauge (in which the remnant gauge symmetry is spontaneously broken)
is necessarily non-confining\cite{zw}. When configurations from a 
standard MC simulation
were transformed to the minimal Coulomb gauge and 4th-dimension pointing links were measured, 
a transition to a ferromagnetic phase was seen\cite{me1}.  It has 
since been found that the gauge-fixing algorithm used in that study was not performing well at 
high $\beta$, which led to too low an 
estimate of the critical $\beta$ of around 2.6.
A more recent study, still underway, with a much improved gauge-fixing algorithm is 
showing this transition to
be around $\beta=3.2$ on the infinite 4-d lattice, with consistent finite size scaling\cite{me2}.  

Regardless of whether this transition occurs at $\beta \simeq 3$, or the standard assumption that
it occurs at  $\beta = \infty$, 
it is interesting to consider the nature of the non-ferromagnetic phase.  
There are two possibilities.  The higher pseudotemperature associated with lowering $\beta$ could induce a 
transition directly to a paramagnetic phase, or there could first be a transition to a spin-glass phase induced
by the higher interaction disorder, followed
by a second thermally-induced transition to the paramagnetic phase. Due to 
the complete randomness of the strong coupling 
limit, the system {\em must} eventually enter a paramagnetic phase.  It is this second scenario, in which
the intermediate-coupling phase is a spin glass, that is being tested here.  

The nested MC with two real replicas described above is utilized. For the inner MC, where the 
sideways links are fixed and treated as part of the interaction Hamiltonian of the fourth-direction spins, the
gauge symmetry is reduced to the global in three dimensions, local in one, center symmetry, Z2, operating on 
perpendicular hyperlayers.  Such an operation leaves the fixed sideways links, and therefore the interaction
Hamiltonian invariant. It also flips all 4th-direction pointing spins in adjacent hyperlayers. 
To form the spin-glass magnetization order parameter, 
a scalar product is taken between the corresponding
spins of the two
replicas, and these values are summed over the three spatial directions. 
\[ q= \sum _i \vec{s^{a}_{i}}\cdot \vec{s^{b}_{i}}\]
After suitable equilibration, the absolute value of $q$ and its moments are measured.  These
are further averaged over hyperlayers and over a number of 
different ``disorder environments" provided by the conventional
outer MC.  Once the magnetization is obtained the analysis is also conventional, similar to what one
would do for the 3-d Ising Model.  
The order parameter is sensitive to the residual Z2 gauge symmetry described above, if only one of the two replicas 
is transformed. When this symmetry is spontaneously broken the resulting magnetization will be nonzero. In this way
the two-replica method allows one to see states which are frozen, but do not show long-range order in space. Such
systems nevertheless have long range order in MC time and break ergodicity on the infinite lattice.  

\section{Monte Carlo Study}
For this study, the outer MC generated 100 configurations, each of which was separated by 100 sweeps,  
after an initial equilibration of 10,000 sweeps. Each sweep referred to  actually consisted of a pair of sweeps, 
the first one using the 
heat-bath algorithm\cite{creutz1} and the second the overrelaxation algorithm\cite{creutz2}. 
Each inner MC consisted of 10000 equilibration sweeps 
followed by 5000 measurement sweeps. This
resulted in 500,000$\times L$ total measurements (where $L$ is the linear lattice size)
of the spin-glass magnetization and its moments.  Correlation 
functions were also measured
on each configuration, from which the second moment correlation length(SMCL)\cite{smcl} was calculated. Lattices
of size $16^4$, $20^4$, and $24^4$ were measured for a range of $\beta$. 

It is crucial in these kinds of simulations to 
equilibrate long enough in the inner MC.  This is needed to decorrelate the two replicas, which, of course,
start out equal.  In the paramagnetic phase they will decorrelate completely, and in a spin-glass phase they will stay
correlated, as measured by the spin-glass order parameter, forever.  Of course tunnelings between the two broken vacua,
will take place on a finite lattice, so in practice one must measure the absolute value of the
magnetization as usual. A preliminary study was performed at several $\beta$ with different amounts of 
equilibration to determine how 
much was necessary.  An exponential relaxation in MC time was observed for $|q|$ with time constant of 300-1500 sweeps
on the $24^4$ lattice in the range of $\beta$  studied. The relaxation was slowest at $\beta = 1.96$, consistent
with critical slowing down in this region. Fig.~1 shows the results of this 
preliminary equilibration
study, in which lattices were equilibrated for $n$ sweeps followed by 200 measurement sweeps (the outer MC was the same as 
in the main
study described above). Other quantities such as the SMCL behaved similarly. 
Another set of runs employed random starts for the spawned replicas.  Here the 
spin-glass magnetization builds
from zero and appears to be approaching the same equilibrium states as the ordered starts. The random start equilibration 
is quite a lot
slower in the spontaneously broken region, probably because different pieces of the lattice initially 
fall into different broken 
vacua, which then need to be reconciled by tunneling.  Whereas the ordered starts fit nicely to single exponential decay 
for $n > 600$, the random starts, with their long tails, do not.  
From the fits to the ordered start runs, it was
determined that 5000 equilibration sweeps were sufficient to bring the remaining systematic error below 50\% of the 
random error. However, it was decided to double this for a safety factor, since not every $\beta$ was checked. 
With 10,000 equilibrium sweeps and 5000 measurement sweeps as used in the main study, the expected systematic error from
equilibration-effects is projected to less than 2\% of the random error.

As a matter of
interest, $\beta=2.5$, a far weaker coupling but more in the range of usual SU(2) simulations was also checked. 
Here, no effect of inadequate equilibration, using the above parameters, could be seen at all. By reducing the number of 
measurement sweeps to 10, an effect was finally seen, with all spin-glass quantities 
apparently fully equilibrated at this $\beta$ in only 20
sweeps. 
\begin{figure}
\includegraphics[width=3in]{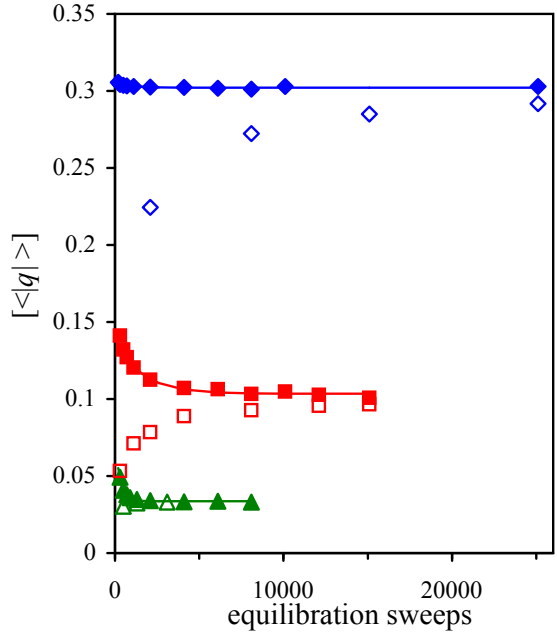}
                                 \caption{Spin-glass magnetization vs. equilibration time. 
Error bars are about $\frac{1}{3}$ symbol size. Diamonds are $\beta = 2.05$, squares $\beta = 1.96$, and triangles 
$\beta = 1.9$. Filled symbols represent ordered starts and open symbols are random starts. Single-exponential fits are 
shown for the ordered starts.}
          \label{fig1}
\end{figure}
\begin{figure}
                      \includegraphics[width=3in]{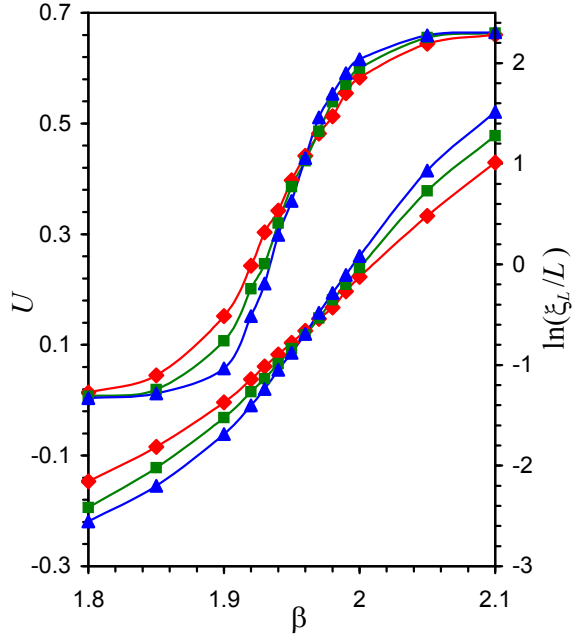}
                                 \caption{Crossings for Binder Cumulant (left) and second moment correlation 
length. Error bars are about $\frac{1}{2}$ symbol size for $\beta < 1.94$ and much smaller for high $\beta$. Crossings 
are verified at 
$\beta = 2.0$ at 15 and 20 standard deviations respectively. Diamonds are 16$^4$, squares 20$^4$, 
and triangles 24$^4$ lattices on this and all subsequent graphs.}
          \label{fig2}
       \end{figure}

The results of the full study, which ranged from $\beta$ = 1.8 to 2.1, show a definite spin-glass 
magnetization in the entire $\beta$-range normally used in SU(2) simulations.
The Binder cumulant, $U=1- [<\! q^4 \! >]/3[<\! q^2 \! >]^2$, for instance, is an increasing function of lattice 
size for $\beta > 1.96$ (Fig.~2), indicating
that this region is a spin-glass magnetized phase on the infinite lattice. Both $U$ and the 
second moment correlation length
divided by lattice size, $\xi _L /L$,  show
clear crossings at  $\beta=1.96 \pm 0.01$ (Fig.~2), indicating an infinite-lattice critical point 
here. Because $\xi _L$ is determined from the unsubtracted correlation function, it diverges not only at the
critical point, but even more strongly in the entire symmetry-broken phase, which is responsible for the crossing behavior.
\begin{figure}
                      \includegraphics[width=3in]{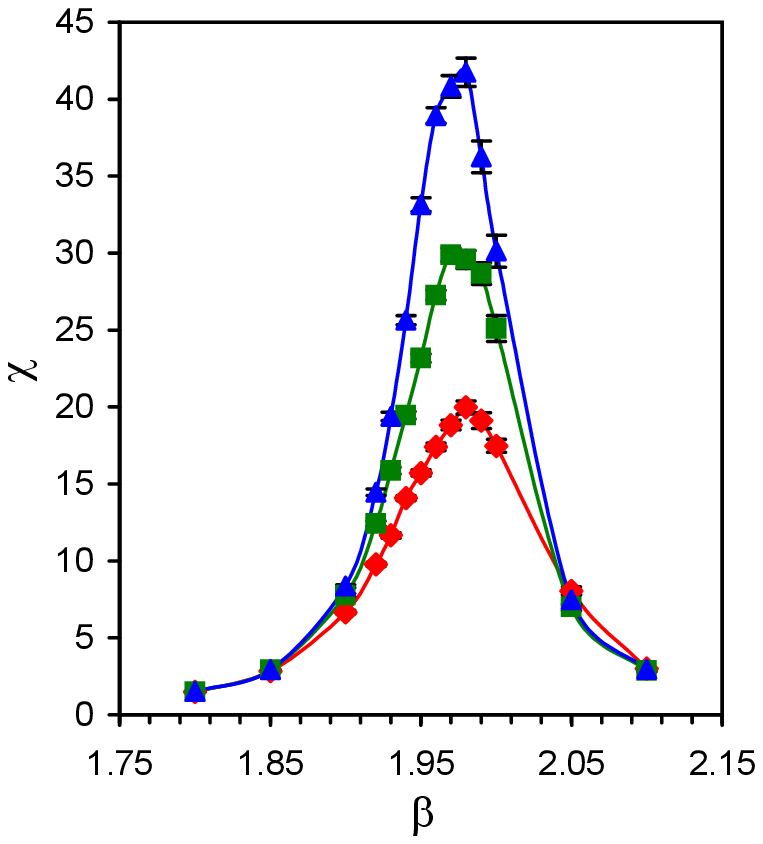}\includegraphics[width=3in]{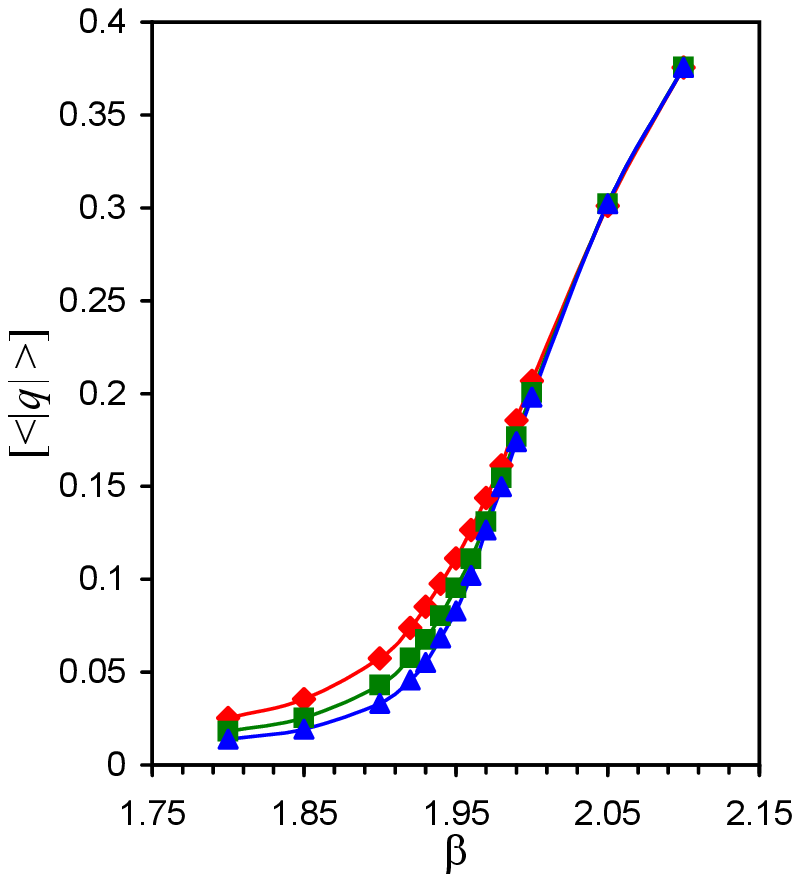}
                                  \caption{(a) Spin-glass susceptibility  and (b) spin-glass magnetization vs. $\beta$.
Error bars for magnetization are about $\frac{1}{3}$ symbol size.}
          \label{fig3}
       \end{figure}
The spin-glass susceptibility, $\chi = [<\! q^2\! >] - [<\! |q|\! >]^2$, shows a growing peak in this region (Fig.~3a).
The spin-glass magnetization itself is shown in Fig.~3b.
Scaling ``collapse plots" were used to get a simultaneous fit to $[<\! |q|\! >]$, $U$, $\chi$, and $\xi _L/L$
on the three lattices. Conventional finite-size
scaling ans\"{a}tze \cite{fss} were used.
First, a nonlinear optimizer
was used to determine the exponents $\gamma /\nu $ and $\beta /\nu $ from collapse plots of $\chi L^{-\gamma /\nu}$
and $[<\! |q|\! >]L^{\beta /\nu}$ vs. $U$.  Plots using U itself as the scaling variable have the advantage that they do
not require determination of $\beta _c$ or $\nu$; each has only one adjustable parameter. These gave results very consistent
with the hyperscaling relationship
$2\beta /\nu + \gamma/\nu =d=3$. Therefore it was decided to enforce that relationship to reduce the parameter count by one.
These collapse plot fits, shown in Figs.~4a and 4b, give $\gamma /\nu = 1.95 \pm 0.16$
and therefore $\beta /\nu = 0.53 \pm 0.08$.  Points from $\beta = 1.9$ through 2.05 were included in the fit, which 
had $\chi ^2/$d.f.$ = 0.8$. Various functions were used for the fits themselves.  Their exact functional forms are irrelevant in this context -
the important thing is that data from the different lattice sizes fall on a single curve, which only happens for this range
of $\gamma /\nu$.
Then the four scaling plots for $[<\! |q|\! >]$, $\chi$, $U$,
and $\xi _L /L$ were performed (Figs.~5a,b) and values of $\beta _c$ and $\nu$ were determined 
by searching for the best collapse. Scaling variables are shown in axes labels. Note $T = 1/\beta$.
These gave $\beta _c = 1.962 \pm 0.006$ and $\nu = 0.99 \pm 0.13$. 
The overall collapse fit for the four quantities on three lattices 
had a $\chi ^2 /$d.f. of 1.6.
Error bars for the critical exponents were obtained by forcing them
higher or lower (allowing other parameters to change) 
until the fit gave a $\chi^2 / d.f.$ of one greater, or double the original, if the original exceeded unity.  This 
more-conservative criterion guards against the possibility that error bars in the data might be slightly 
underestimated. 
Data error bars were determined from the asymptotic behavior of binned fluctuations. Finally, probability
distributions are shown in Fig.~6 for various values of $\beta$ below and above the transition.  These appear to follow the
normal pattern of a higher-order transition.

\begin{figure}
                      \includegraphics[width=3in]{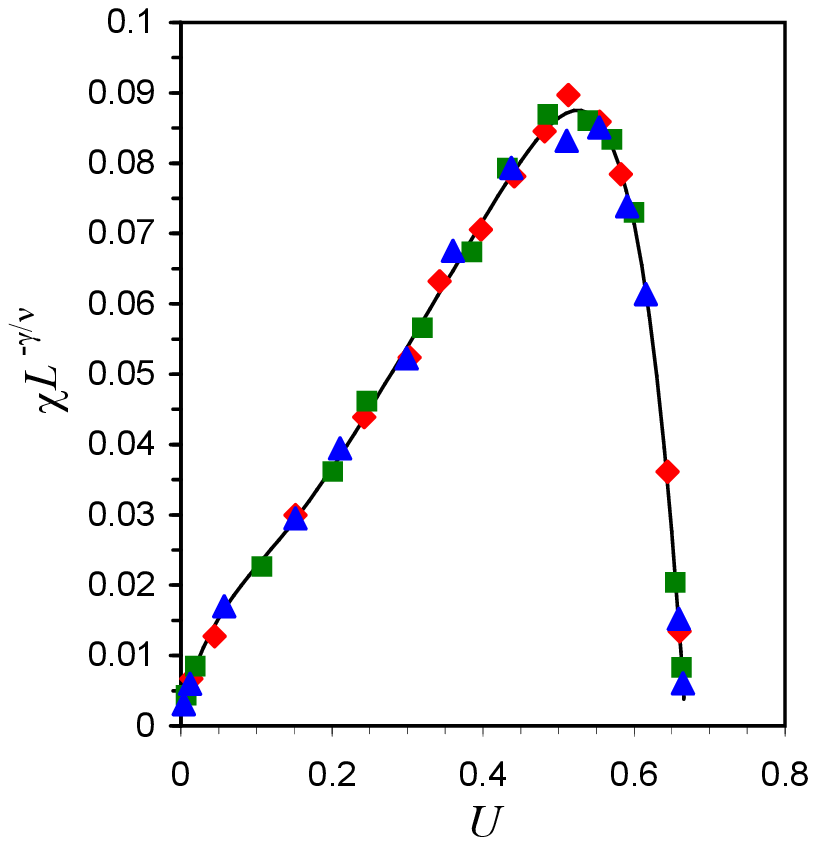}\includegraphics[width=3in]{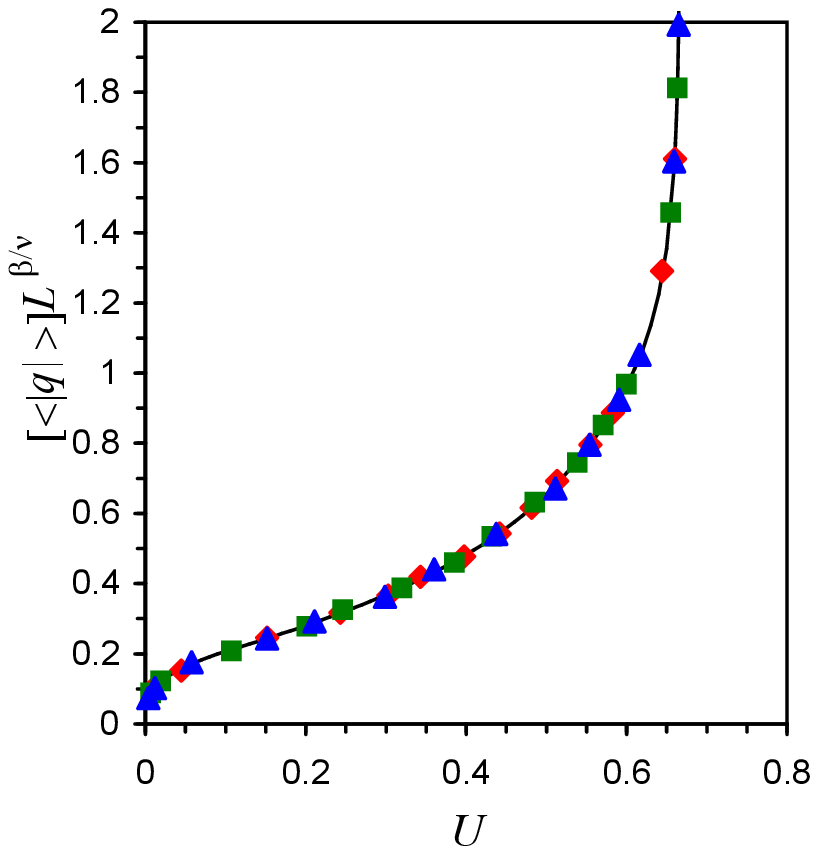}
                               \caption{Scaling collapse plots for (a) spin-glass susceptibility 
and (b) spin-glass magnetization vs. fourth order cumulant, $U$. Error bars are of order symbol size for susceptibility,
and about $\frac{1}{3}$ symbol size for magnetization. Unlike the other plots, these also 
have horizontal uncertainties of order 
$\frac{1}{2}$ symbol size for $U < 0.5$ and much smaller above this.}
          \label{fig4}
       \end{figure}
\begin{figure}
                      \includegraphics[width=3in]{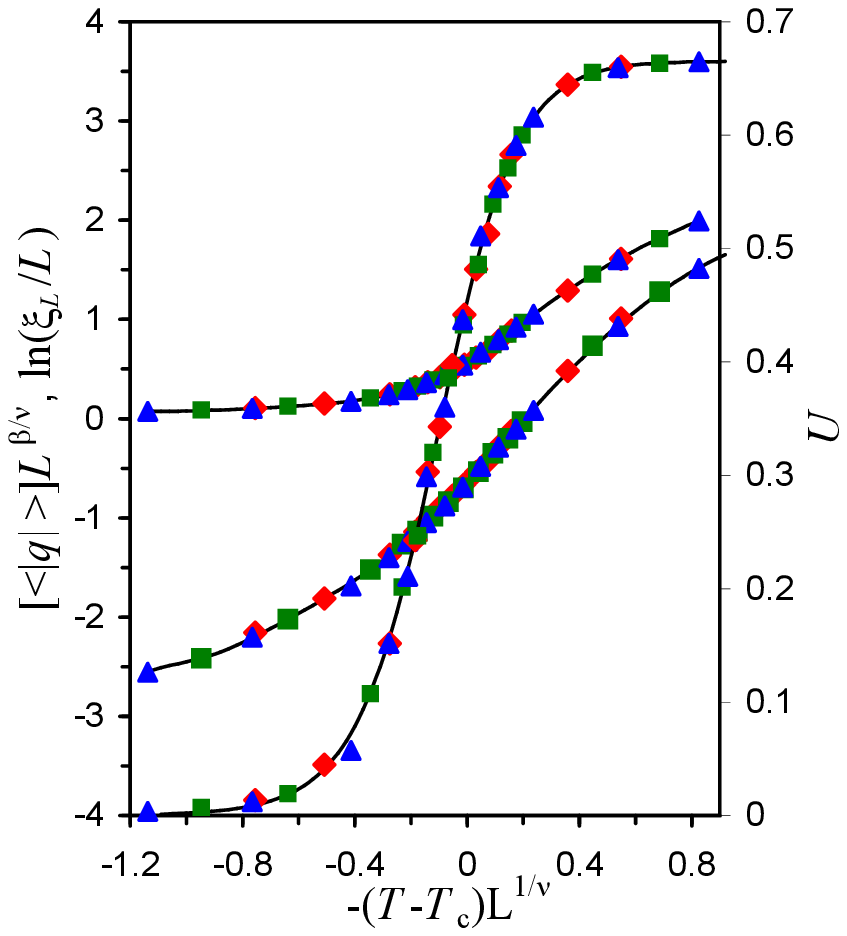}\includegraphics[width=3in]{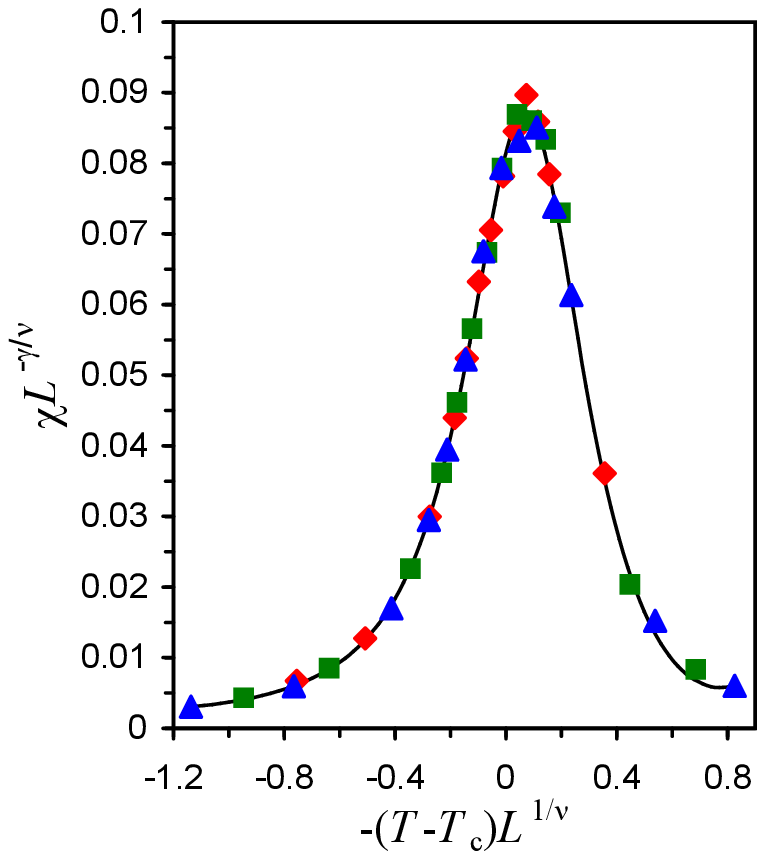}
                               \caption{Scaling collapse plots for (a) spin-glass magnetization 
(left), $\xi _L/L$, and $U$; (b) Collapse plot for spin-glass susceptibility.
Error bars for magnetization are about $\frac{1}{2}$ symbol size, and for susceptibility, 
the full symbol (for other errors see Fig.~2).}
          \label{fig5}
       \end{figure}

A relatively large value of $\nu$ is typical of 
spin glasses. For instance for the 4-d $\pm J$ Ising spin glass 
an MC study gives $\nu = 1.0 \pm 0.1 $\cite{isg}, and for the O(3) 3-d Heisenberg $\pm J$ 
the value $1.04 \pm 0.06$ has been reported\cite{hsg}.  
Large values of $\nu$ all
but erase a detectable signal of a phase transition in the specific heat, which initially led to the questioning of whether
the spin-glass transition was a true phase transition. Phase transitions were affirmed by studying other quantities such
as the correlation length. 

\begin{figure}
                      \includegraphics[width=3in]{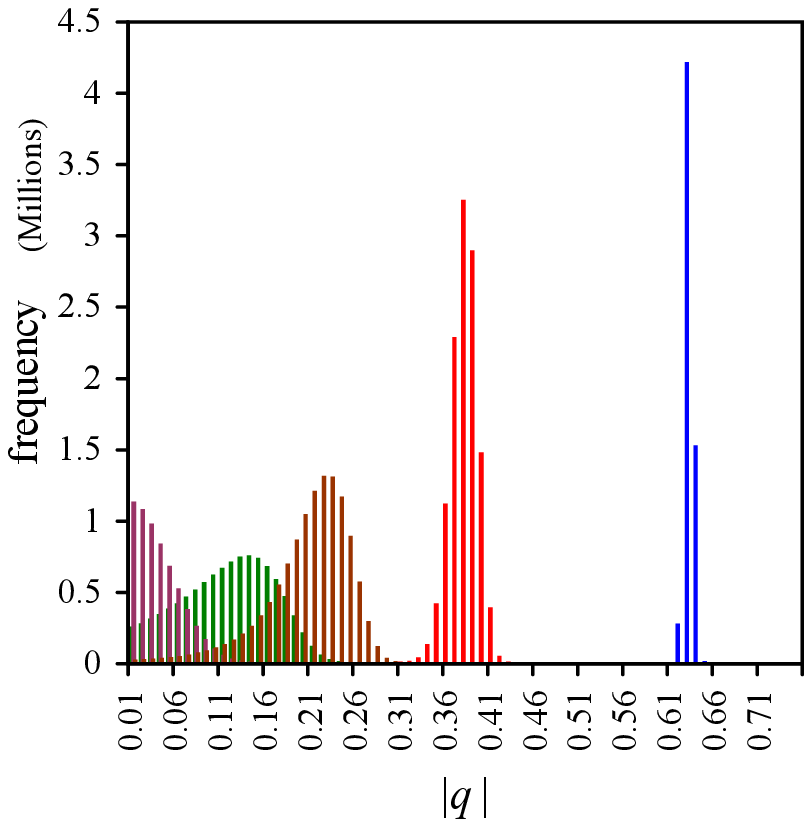}
                               \caption{Probability distributions for spin-glass magnetization for 
$\beta=$1.9, 1.96, 2.0, 2.1, and 2.5}
          \label{fig6}
       \end{figure}

\section{Roughening Transition}
The critical point of 1.96 is close to that of the previously-known roughening transition which is most often 
estimated at ``around" $\beta _R = 1.9$, from an apparent developing singularity in the 
strong-coupling expansion\cite{roughening}.  Different Pade's give varying estimates 
giving an uncertainty in  $\beta _R$ of around 0.1 . 
Thus it is possible the two transitions coincide.  Connections between roughening and spin 
glasses have been suggested in other 
contexts\cite{sg-rough}. Roughening, however, is expected to fall in the Berezinskii-Kosterlitz-Thouless(BKT) class, 
whereas the
transition here is clearly not. For instance in BKT the Binder cumulants of different lattice sizes
would be expected to merge in the weaker-coupling phase, rather than cross, and there is no spontaneous magnetization
in BKT, whereas here the clear $U$-crossing is a strong indication
that nonzero spin-glass magnetization  will survive 
the infinite lattice limit. Also, the order parameter distribution (Fig.~6) has no support at the 
origin for $\beta \geq 2.05$, which
is also inconsistent with BKT scaling.
Whereas the physics of roughening is clear in three dimensions, it
is not as well understood in four; it seems possible the critical behavior could be different than expected. Another 
possibility is that the strong coupling expansion singularity is due to something other than 2-d roughening.
These possibilities require further investigation.  Another important point is that, unlike the roughening transition, 
a spin-glass transition is a true thermal transition, albeit a weak one.  In other words it does induce a singularity
in the free energy on the infinite lattice. This shows that the strong and weak 
limits of the SU(2) LGT are not analytically connected, as is usually assumed.  Both the paramagnetic and
spin-glass phases are, however, confining.  

\section{Conclusion}
The data given above show that the SU(2) lattice gauge theory exhibits 
spin-glass order for all $\beta > 1.96$. 
Since all spin glasses eventually have a second transition to a ferromagnetic phase as the interaction-disorder is decreased,
this would appear to strengthen the case for such a second transition, especially since at $\beta = \infty$ the theory
in the minimal Coulomb gauge becomes a non-interacting set of 3-d O(4) Heisenberg models at zero temperature, 
which are definitely
ferromagnetic.  The question of whether the continuum limit is confining boils down to whether or not 
ferromagnetism is robust as one enters from the edge of the phase diagram, i.e. as $\beta$ becomes finite.  
This would seem to open the possibility of a new
route to proving or disproving confinement, perhaps even analytically.  The fact that it is a spin glass to ferromagnetic
transition would explain why it was not noticed previously.  The high value of $\nu$ typical in 
spin-glass transitions (both to paramagnetic and ferromagnetic states) results
in almost no discernable signal in the specific heat.  
New evidence for such a transition around $\beta = 3.2$ will be detailed soon.
However, regardless of whether this second transition is at finite or infinite $\beta$, the observation of spin-glass 
order in SU(2) lattice gauge theory is of interest in itself, and
may result in new insights into the 4-d roughening transition and the 
lattice QCD vacuum.

\end{document}